\documentclass[reprint,eqsecnum,floats,aps,amsmath,amssymb,nofootinbib,prd,onecolumn,showpacs,superscriptaddress]{revtex4-2}

\usepackage{graphicx}
\usepackage{bm}
\usepackage{amsmath}	
\usepackage{amsfonts}
\usepackage{mathtools}
\usepackage{amssymb}							
\usepackage{braket}
\usepackage{amsthm}			
\usepackage{hyperref}							

\usepackage[normalem]{ulem}
\usepackage[T1]{fontenc} 

\usepackage{color}
\usepackage[dvipsnames]{xcolor}

\usepackage{nomencl}
\makenomenclature

\begin{document}

\title{Robustness of the primordial power spectrum in hybrid loop quantum cosmology to approximations near the bounce}

\author{Kristina Giesel}
\email{kristina.giesel@fau.de}
\affiliation{Institute for Quantum Gravity, Theoretical Physics III, Department of Physics,  Friedrich-Alexander-Universit\"at Erlangen-N\"urnberg, Staudtstr. 7, 91058 Erlangen, Germany}

\author{Almudena Guill\'en}
\email{almudena.guillen@iem.cfmac.csic.es}
\affiliation{Instituto de Estructura de la Materia, IEM-CSIC, Serrano 121, 28006 Madrid, Spain}

\author{Guillermo A. Mena Marug\'an}
\email{mena@iem.cfmac.csic.es}
\affiliation{Instituto de Estructura de la Materia, IEM-CSIC, Serrano 121, 28006 Madrid, Spain}

\author{Leonardo Ricci}
\email{ricci.1968073@studenti.uniroma1.it}
\affiliation{Institute for Quantum Gravity, Theoretical Physics III, Department of Physics,  Friedrich-Alexander-Universit\"at Erlangen-N\"urnberg, Staudtstr. 7, 91058 Erlangen, Germany}
\affiliation{Dipartimento di Fisica, Università degli Studi di Roma ``La Sapienza'', Piazzale Aldo Moro 5, I-00185 Rome, Italy}

\begin{abstract}
We prove the robustness of the analytic approximation used in (hybrid) loop quantum cosmology to compute the primordial power spectrum of the cosmological perturbations for a recently proposed vacuum state (the non-oscillatory state with asymptotic Hamiltonian diagonalization). To investigate this, we study different approximations to the effective mass of these perturbations near the bounce, showing that the Pöschl-Teller potential employed in previous works leads to indistinguishable power spectra compared to other estimations of the mass, provided that they successfully capture the characteristic scale of the bounce, which translates into a scale of power suppression. 
\end{abstract}

\maketitle

\section{Introduction}
\label{Sec:Intro}

The cosmic microwave background (CMB) \cite{HD,Planck_parameters} is an important observable in cosmology. It acts as a window to the primitive Universe, allowing the comparison of predictions from theoretical models of the early evolution with late-time observables \cite{WMAP,Planck_parameters,Planck_Constraints,cutoff1,Planck_anomalies}. Observational data fit in good agreement with a phase of slow-roll inflation driven by a single inflaton field \cite{Planck_Constraints,cutoff1}. However, it has been reported that there are certain statistical anomalies in the observations compared to the predictions of standard $\Lambda$CDM cosmology \cite{Planck_anomalies} (where $\Lambda$ stands for the cosmological constant and CDM for cold dark matter). One of these anomalies is the suppression of power at low multipoles in the angular power spectrum of temperature anisotropies \cite{Linde,cutoff2,cutoff3}. In addition, an important conceptual challenge in connecting theory with observations is the large mismatch between the energy-density scale of slow-roll inflation and that of the Planck-dominated regime \cite{Liu}, at least for mechanisms that assume cosmological perturbations ultimately originate from quantum gravity effects.

Among the proposals to address these problems, quantum cosmology scenarios have emerged as interesting candidates, since they can provide a mechanism to explain phenomena such as, for example, the suppression at large angular scales, owing to curvature effects in the Planck era. In particular, effective models in loop quantum cosmology (LQC) \cite{LQC,Math_LQC} have raised significant interest in recent years, supporting the lack of power at these scales in their predictions of the primordial power spectrum (PPS) \cite{ASr,ASr2,AgSr2}.  

LQC provides a quantum framework for the consideration of cosmological spacetimes based on techniques inherited from loop quantum gravity (LQG) \cite{LQG,Thie}. One of its most important predictions is the replacement of the classical big-bang singularity by a quantum bounce \cite{APS,APS1,Further}, caused by genuinely quantum-geometric effects that become relevant in the Planck regime. Effective descriptions of LQC reproduce classical general relativity (GR) at low energy densities, while introducing characteristic modifications when approaching the Planck density scale near the bounce \cite{APS,BCL,NBMmass}.

In these models \cite{NM,MVY,MVY2}, there is a preinflationary evolution in which the Universe expands from the bounce. Near this bounce, quantum geometry effects are relevant. Furthermore, for effective solutions of phenomenological interest in LQC, for which the imprints of quantum gravity phenomena are neither inadmissibly large nor ignorable nowadays, the kinetic energy density of the inflaton is much bigger than its potential at this early stage \cite{AM}. With the expansion, this energy density decreases and LQC effects become negligible, but the kinetic contribution still dominates over the potential in the inflaton dynamics. In this way, one reaches a classical phase of kinetic domination, a regime compatible with observational constraints \cite{KM}. Eventually, the potential contribution finally surpasses its kinetic counterpart, giving rise to an inflationary period. For reasons of phenomenological interest, again, one considers solutions in which the quantum geometry imprints are not totally diluted during this inflationary period. This leads to the consideration of scenarios of just-enough inflation \cite{Enough_inf}. For simplicity and manageability, their dynamics is usually described in a slow-roll approximation. 

There exist two main strategies to quantize perturbations in LQC while respecting the ultraviolet behavior of the $\Lambda$CDM model and its dispersion relations in this limit: the dressed metric \cite{dressed1,dressed2,AAN,AM} and the hybrid  \cite{hybrid_rev,hybr_inf1,hybr_inf2,hybr_ten,GMdBO} approaches. The former describes perturbations propagating on an effective quantum-corrected geometry, whereas the latter is based on a canonical quantization of the perturbed cosmological system (truncated at quadratic perturbative order in the action). In both cases, the quantization is carried out by combining LQG inspired and Fock techniques. Although both approaches agree in the classical regime, they generally predict different effective equations for the perturbations in the high-curvature epoch. In the solutions of phenomenological interest under consideration, only the largest observable modes felt the quantum curvature near the bounce, leading to a suppression of power in that sector. For both approaches, suppression of power has been reported in several works, with differences in the amount and the scale of the suppression \cite{GLMRV}. 

Focusing on analytical studies of this kind to obtain predictions for the PPS, several works in effective LQC have carried out the following procedure (this includes, for instance, Refs. \cite{NM,MVY2,GLMRV}). First, the background evolution is divided into (at least) the three epochs previously mentioned, in which it is possible to analytically solve the dynamics after certain approximations: bounce epoch, classical kinetic domination, and slow-roll inflation. Instantaneous transitions are generally assumed for the sake of simplicity. For the perturbations, their effective mass in the mode propagation equations of the modified Mukhanov-Sasaki equation is commonly approximated in the vicinity of the bounce by a Pöschl-Teller (PT) potential (see e.g. Refs. \cite{PT-1,waco2,NM}). This mass behaves like in GR in the kinetic domain, and finally becomes the standard slow-roll mass term during inflation, allowing in both cases an analytic description. Continuity of the mode solutions to the propagation equations of the perturbations is applied up to the first derivative, as the field equations are second order. So, once the initial conditions of the perturbations are specified for any of the three eras, the evolution is uniquely determined for the whole preinflationary and inflationary periods \cite{NM}.

In this framework, the approximated effective mass of the perturbations used in the computations is not smooth, and does not even need to be continuous at the matching points between the different epochs (for example, continuity is generically lost at the matching between the bounce era and the kinetically dominated regime). Following the discussion in Ref. \cite{NM}, this non-smoothness leads to the appearance of spurious fast oscillations in the mode solutions and, therefore, in their PPS, a phenomenon that can be regarded as an artificial enhancement of power. However, it is possible to get rid of them by applying a Bogoliubov transformation \cite{NM}. The resulting non-oscillating power spectrum displays suppression at small wavenumbers, determined by specific parameters of the model.

A highly relevant ingredient in the calculation of the power spectrum is the vacuum state selected for the primordial perturbations. When a preinflationary evolution is considered, the choice of the Bunch-Davies \cite{Bunch,Mukhanov,WaldH} state as the preferred vacuum loses a solid justification, as discussed e.g. in Refs. \cite{NM,KM,NMT,NdBM}. A vacuum should be a state optimally adapted to the properties and dynamics of the background, and the Bunch-Davies state only satisfies this requirement in quasi-de Sitter regimes. Modes capable of feeling departures from this regime should not be expected to be in the Bunch-Davies solution at the end of inflation. It is therefore crucial to develop a general criterion to select a vacuum state in these situations with a quantum preinflationary dynamics, preferably based on fundamental physical considerations. In this work, we will adhere to the so-called NO-AHD proposal \cite{NMT}, which selects positive-frequency solutions in the asymptotic regime of large wavenumbers and displays non-oscillating (NO) properties. In consonance with our considerations, it is remarkably well adapted to the background dynamics, making it very suitable for capturing quantum effects arising from the preinflationary era. Moreover, it is one of the proposals for which suppression of power in the infrared region has been reported \cite{NM,MVY,MVY2,GLMRV}. However, in the analytic studies considered in LQC, given the lack of smoothness in the effective mass of the perturbations at the matching points, the vacuum obtained by asymptotic Hamiltonian diagonalization (AHD) during the bounce epoch would not be perfectly adapted to the whole preinflationary and inflationary evolution \cite{NM}. The problem would be the development of fast oscillations. Nevertheless, as we have already mentioned, following Ref. \cite{NM}, we can remove them at the end of inflation by a Bogoliubov transformation, achieving in this way a good NO-AHD vacuum for the perturbations.
 
Our goal in this work is to check whether the PPS predicted by the NO-AHD vacuum depends critically or not on the approximation used for the effective mass of the perturbations in the quantum era. In fact, the approximations in the other, classical epochs have already been checked by comparing exact numerical calculations with approximate analytic computations, proving the extraordinary accuracy of the analytic treatment \cite{MVY}. This leaves the approximation employed near the bounce as the only still questionable component. To check its validity, we compare the PT approximation to the mass with three other different models, respectively obtained by approximating the effective mass by a double PT potential, by a Rosen-Morse (RM) potential, or by the mass term of a de Sitter (dS) expanding cosmology. Note that this last model implies also a different background evolution near the bounce.

For conciseness, we restrict ourselves to the discussion of scalar perturbations and consider exclusively the hybrid approach to LQC (except for the dS case). We maintain the study analytical wherever possible. Given the similarity in the methodology and results for the PPS obtained in previous works for tensor perturbations, as well as by adopting the dressed metric approach, we expect similar conclusions in those cases, as we will briefly comment in the last section. 

The content of the rest of the paper is organized as follows. In Sec. \ref{Sec:Initial_framework}, we summarize the background evolution and dynamics of perturbations for a PT effective mass and the NO-AHD vacuum in hybrid LQC. In Sec. \ref{Sec:Alternative_mass_terms}, we present the alternative approximations of the effective mass for the quantum era and explain their main features. Section \ref{Sec:PPS} is devoted to the study of the PPS. Finally, in Sec. \ref{Sec:Discussion} we discuss our results and conclude. Throughout the article we work in Planck units, with $\hbar = G = c = 1$.

\section{The model with Pöschl-Teller potential}
\label{Sec:Initial_framework}

In this section, we briefly review the dynamics of the background and the linear perturbations in hybrid LQC, imposing the NO-AHD criterion to select the initial conditions for the perturbations and approximating their effective mass as a PT potential near the quantum bounce.

\subsection{Background evolution}
\label{Subsec:Background_evolution}

As we have commented, we distinguish three stages in the evolution of the cosmological background. The first of these corresponds to the vicinity of the bounce, in which the dynamics is sensitive to the quantum geometry effects of LQC. The modified Friedmann equations are given by \cite{AAN,GMdBO}

\begin{equation}\label{eq_Friedman_LQC}
 \left( \frac{a'}{a}\right)^{2} = \frac{8\pi}{3}a^{2}\rho \left(1 - \frac{\rho}{\rho_{c}} \right),\qquad \frac{a''}{a} = \frac{4\pi}{3}a^{2}\rho \left(1 + 2\frac{\rho}{\rho_{c}} \right) - 4\pi a^{2}P \left(1 - 2\frac{\rho}{\rho_{c}} \right).
\end{equation}
Here, $a$ is the scale factor, the prime denotes the derivative with respect to the conformal time $\eta$ and $\rho = \left (\phi'/a \right )^2/2 + V$ and $P = \rho - 2V$ are, respectively, the energy density and pressure of the homogeneous inflaton field $\phi$ with potential $V := V(\phi)$. On the other hand, $\rho_c = 3/(8\pi \gamma^2\Delta)$ is the critical energy density in LQC \cite{LQC}, where  $\Delta=4\sqrt{3}\pi \gamma$ is the area gap in LQG \cite{LQG,Thie} and $\gamma$ is the Immirzi parameter \cite{immirzi}. Among the different regularizations in LQC \cite{Yang:2009fp}, we adopt the standard value of $\rho_c$ for $\gamma=0.2375$, motivated by black-hole entropy calculations in LQG \cite{LQG}. This choice has been widely used in several studies of the PPS in LQC (e.g. \cite{GLMRV,ASr,NM,GMdBO}), and therefore allows for better comparison with the one presented here. 

In the following, for concreteness in our calculations, we assume a quadratic potential for the inflaton of the form $V(\phi)=m^2\phi^2/2$, with a value of $m=1.2 \times 10^{-6}$ in Planck units, in line with previous works in the literature \cite{AM,NM,GLMRV}. 

It can be shown that the contribution of this potential to the energy density of the inflaton is practically negligible until later times in the evolution in the considered solutions of phenomenological interest \cite{waco2,NMY}, so we can set $V(\phi)=0$ in Eqs \eqref{eq_Friedman_LQC} around the bounce. Thus, the dynamics are analytically solvable in proper time $t$, yielding \cite{NM,waco2}
\begin{equation}\label{scale_factor_LQC}
 a(t) = \left (1 + 24\pi \rho_c t^2 \right )^{1/6}, \qquad \dot{\phi }(t)=\pm \sqrt{2\rho_c}a^{-3}(t).
\end{equation}
The sign in the second identity corresponds to the two possibilities of increasing or decreasing inflaton field at the bounce, where we take a value of $\phi_0=0.97$ for our calculations, which is also a frequent choice for a quadratic potential in the literature \cite{NM,MVY,MVY2,GLMRV}. Furthermore, note that we have set $t=0$ at the bounce and $a(0)=a_B=1$. Therefore, when comparing results to observational data, for which one adopts the convention that the present scale factor equals the unit, we will have to scale all length observables proportionally to the ratio of scale factors at the bounce and today. On the other hand, the relation between proper and conformal times in the quantum era is  

\begin{equation}\label{relation_eta_and_t}
    \eta = {}_{2}F_1\left (1/6, 1/2; 3/2;-24\pi \rho_ct^2 \right )t,
\end{equation}
where ${}_{2}F_1$ is the Gauss hypergeometric function \cite{Abra} and we have set to zero the conformal time at the bounce.

Once the energy density decreases sufficiently, the LQC solutions become indistinguishable from those of a classical period of kinetic domination. We consider the transition to be instantaneous, at a time $\eta_0$ or equivalently $t_0$. For concreteness, we will adopt a value of $t_0=0.41$, which has been proved to be a good estimation for the quadratic inflaton potential \cite{NM} and has been used in several works \cite{NM,MVY,MVY2,GLMRV,NMY}.  

During classical kinetic domination, background equations are those given in Eq. \eqref{eq_Friedman_LQC} disregarding the terms $\rho/\rho_c$. It is still valid to ignore any effect of the inflaton potential on the background solutions, which therefore take the form 

\begin{equation}\label{scale_factor_KD}
 a(\eta) = a_0\sqrt{1 + 2a_0 H_0 (\eta - \eta_0)} \, , \qquad H(\eta)= H_0 \left (\frac{a_0}{a(\eta)} \right)^3.
\end{equation}
Here, $H$ is the Hubble parameter in cosmic time, $H=\sqrt{8\pi \rho/3}$, although we have expressed it as a function of $\eta$. Initial conditions $H_0=H(\eta_0)$ and $a_0=a(\eta_0)$ are obtained by evaluating the scale factor and the Hubble parameter from the quantum era at the matching point $\eta_0$.   

The contribution of the inflation potential eventually becomes comparable to its kinetic counterpart, resulting in a period of inflation, which we describe in a slow-roll regime. The slow-roll parameters are given by \cite{Langlois,Baumann}

\begin{equation}\label{sr_parameters}
\epsilon_V = \frac{1}{16\pi}  \left ( \frac{V,_{\phi}(\phi)}{V(\phi)}\right)^2 , \qquad \delta_V = \frac{1}{8\pi} \frac{V,_{\phi \phi}(\phi)}{V(\phi)},
\end{equation}
where the comma denotes the derivative with respect to the inflaton field. We assume that slow-roll lasts until these parameters cease to behave as constants, while remaining much smaller than unity. We denote this instant by $\eta_e$, whereas $\eta_i$ marks the beginning of this inflationary phase \footnote{Numerical simulations provide values of $\eta_i \simeq 912$ and $\eta_e \simeq 2788$ \cite{GLMRV}, and for the slow-roll parameters we take $\epsilon_V\approx \delta_V \simeq 0.0086$.}. Again, initial conditions can be matched to the corresponding values at the end of kinetic domination.

At first order in slow-roll, background dynamics can be computed using the approximation \cite{Baumann}

\begin{equation}\label{equation_sr}
\frac{\mathrm{d}}{\mathrm{d}\eta}\left (\frac{a}{a'} \right) \approx \epsilon_V - 1,      
\end{equation}
which leads to 

\begin{equation}\label{conformal_time_sr}
\eta_e - \eta \approx \frac{1}{aH}(1 + \epsilon_V)
\end{equation}

for any instant within slow-roll inflation.

\subsection{Dynamics of perturbations}
\label{Subsec:Dynamics_of_perturbations}

The evolution of linear scalar perturbations is dictated by the Mukhanov-Sasaki equation \cite{LGomar,NBMmass}

\begin{equation}\label{equation_MS}
v_{\vec{k}}'' + \left (k^2 + s \right )v_{\vec{k}} \, ,  
\end{equation}
where  $v_{\vec{k}}$ is the Fourier coefficient of the Mukhanov-Sasaki (MS) field \cite{Mukhanov,Sasaki,Sasaki1} with wavevector $\vec{k}$, $k = |\vec{k}| $ is the wavenumber, and $s$ plays the role of a time-dependent effective mass for the perturbations. We want to solve this equation for the three epochs presented above\footnote{In all the three epochs, solutions are normalized with respect to the Klein-Gordon product.}, for which the effective mass is different in each case. 

This mass can be analytically computed during the preinflationary and inflationary eras except in the quantum regime near the bounce. For this bounce period in hybrid LQC we denote the effective mass by $s_h$ (the subindex $h$ standing for hybrid). It can be expressed in terms of the cosmological time as \cite{NM}

\begin{equation}\label{LQC_mass}
    s_h(t) = \frac{8 \pi \rho_c}{3} \left (1 + 24\pi \rho_c t^2 \right)^{-2/3},
\end{equation}
but no closed expression in terms of the conformal time is available around the bounce. In order to explicitly solve the MS equation in that interval, it is widely approximated by a PT potential of the form \cite{waco2,waco,NM}

\begin{equation}\label{PT_potential_and_parameters}
s_h \approx s_{PT} = \frac{U_0}{\cosh^2(\alpha \eta )}, \qquad U_0 = \frac{8\pi \rho_c}{3}, \qquad \alpha = \frac{\operatorname{arccosh}(a_0^2)}{\eta_0}.
\end{equation}
Here, $U_0$ and $\alpha$ represent, respectively, the peak and width of the potential. It has been proposed to determine these parameters by imposing that $s_{PT}$ equals $s_h$ at the bounce and $s_{kin}$ (the mass term during the classical era of kinetic domination) at $\eta_0$ \cite{NM}. These requirements have been checked to minimize the relative error between the hybrid LQC mass and the PT potential,

\begin{equation}\label{relative_error}
\text{Err} = 2 \frac{|s_h-s_{PT}|}{(s_h+s_{PT})} \,.   
\end{equation}
For the quadratic potential, this error remains below $0.15$ during the whole quantum era \cite{NM}. 

The mode solutions in the Planck era for the PT potential are given by

\begin{equation}\label{LQC_mode_solutions}
    \mu_k = M_k \left [ x(1 - x) \right ]^{-ik/(2\alpha)} {}_{2}F_1 \left (b_1^k, b_2^k; b_3^k; x \right ) + N_k  \left [\frac{x}{1 - x} \right]^{ik/(2\alpha)} {}_{2}F_1 \left (b_1^k - b_3^k + 1, b_2^k - b_3^k + 1; 2 - b_3^k; x \right ),
\end{equation}
where $M_k$ and $N_k$ are integration constants, $x= \left [1 + e^{-2\alpha\eta} \right ]^{-1}$ is the time variable, and the parameters of the hypergeometric function, which depend on $\alpha$, $U_0$ and $k$, are given by

\begin{equation}\label{parameters_hypergeometric_function}
    b_{1}^k = \frac{1}{2} \left (1 + \sqrt{1 + \frac{4U_0}{\alpha^2}} \right ) - \frac{ik}{\alpha}, \qquad b_{2}^k=b_{1}^k-\sqrt{1 + \frac{4U_0}{\alpha^2}},\qquad b_3^k = 1 - \frac{ik}{\alpha}.
\end{equation}

Next, the mass term during the epoch of kinetic domination reads \cite{NM}

\begin{equation}\label{mass_KD}
    s_{kin} = \frac{1}{4} \left (\eta - \eta_0 + \frac{1}{2a_0 H_0} \right ), \qquad a_0 = \left (1 + 24\pi \rho_c t_0^2 \right )^{1/6} , \qquad H_0 = \frac{\sqrt{U_0}}{a_0^3},
\end{equation}
which yields the following solution to the mode equation \eqref{equation_MS}:

\begin{equation}\label{KD_mode_solutions}
    \mu_k = C_k \sqrt{\frac{\pi y}{4}} H_0^{(1)}(ky) + D_k \sqrt{\frac{\pi y}{4}} H_0^{(2)}(ky), \qquad y = \eta - \eta_0 + \frac{1}{2a_0H_0}.
\end{equation}
Here, $C_k$ and $D_k$ are integration constants, and $H_\nu^{(1)}$ and $H_\nu^{(2)}$ denote the Hankel functions of the first and second kind, respectively \cite{Abra}.

Lastly, during slow-roll inflation the effective mass of the scalar perturbations behaves as \cite{Baumann,NMY}

\begin{equation}\label{mass_sr}
    s_{sr} = - \frac{1}{(\eta - \eta_e)^2} \left (\nu^2 - \frac{1}{4} \right), \qquad \nu = \sqrt{\frac{9}{4} + 9 \epsilon_V - 3\delta_V} \, . 
\end{equation}

The corresponding perturbative mode solutions are

\begin{equation}\label{sr_mode_solutions}
    \mu_k = A_k \sqrt{\frac{\pi }{4}(\eta_e - \eta)} \, H_\nu^{(1)}[k(\eta_e - \eta)] + B_k \sqrt{\frac{\pi}{4}(\eta_e - \eta)} \, H_\nu^{(2)}[k(\eta_e - \eta)],
\end{equation}
with $A_k$ and $B_k$ being integration constants.

Once we have the solution to the perturbative modes for the different epochs of our preinflationary history, to uniquely determine the evolution of the scalar perturbations we must select a vacuum state. Since we want to capture the quantum geometry effects near the bounce, we apply our criterion for the choice of a vacuum state with asymptotic Hamiltonian dynamics in the Planck regime governed by LQC dynamics. Note that, in this regime, we have smooth analytic expressions at our disposal. We then fix the rest of the evolution of the modes by imposing their continuity up to the first derivative at the matching points between consecutive epochs. In this way, we obtain expressions that connect the coefficients of one epoch with those of the next one, so that they all get determined once we specify the constants $M_k$ and $N_k$ by the selection of the vacuum state around the bounce \cite{NM,MVY,MVY2,NMY,GLMRV}. We recall that, in this analysis, the lack of smoothness of the effective mass at the matching points entails that the selected vacuum state at the bounce era will develop fast oscillations, losing its optimal adaptation to the background dynamics  \cite{NM}. However, as discussed above, these oscillations can be removed by a suitable Bogoliubov transformation that finally provides the desired NO-AHD state for the scalar perturbations at the end of inflation. 

Although this NO-AHD prescription was originally developed within the framework of hybrid LQC \cite{NMT}, its construction is sufficiently general to be applied in a wide variety of cosmological settings, including the dressed metric approach to LQC. The NO-AHD vacuum is associated with a family of annihilation and creation variables whose evolution is governed by a perturbative Hamiltonian that becomes asymptotically diagonal in the ultraviolet sector, with the self-interaction terms removed order by order in this regime. Equivalently, it selects a preferred family of positive-frequency solutions that are optimally adapted to the dynamics of the cosmological background. These solutions are given by

\begin{equation}\label{NO-AHD_modes}
   \mu_k = \frac{1}{\sqrt{-2\Im(h_k)}} \, e^{i\int_{\eta_0}^{\eta}d\tilde{\eta}\Im(h_k)(\tilde{\eta})},
\end{equation}
where $\Im$ denotes the imaginary part and $h_k$ is the so-called complex frequency function, obtained as a solution to the Riccati equation $h'_k = k^2 + s + h_k^2$ such that $kh_k^{-1}$ admits an asymptotic ultraviolet expansion as a power series of $k^{-2}$, with leading and subleading terms of the form \cite{NM}

\begin{equation}\label{NO-AHD_equation}
   kh_k^{-1} \sim i\left (1 - \frac{s}{2k^2} \right) . 
\end{equation}
Here, $s$ denotes again the effective mass of the perturbations. In our case, since we apply the criterion in the bounce epoch, it corresponds to the (approximated) mass in hybrid LQC. In the case of the PT potential, this prescription yields $M_k = 1 / \sqrt{2k}$ and $N_k = 0$ for the mode solutions \eqref{LQC_mode_solutions} \cite{NM}.

In light of this construction and the prominent role played in it by the PT approximation, we investigate whether an alternative and different approximation to the effective mass of the perturbations in the vicinity of the bounce will significantly affect the NO-AHD state and therefore the PPS. We recall that the validity of our approximations during the later classical stages has already been extensively checked in the literature. In the following, therefore, we concentrate our attention on the bounce period, discussing the sensitivity of the primordial power spectrum to the approximation employed for the effective mass in this epoch.

\section{Alternative approximations of the effective mass term in the quantum era}
\label{Sec:Alternative_mass_terms}

\begin{figure}[h]
    \centering
    \includegraphics[width=0.7\textwidth]{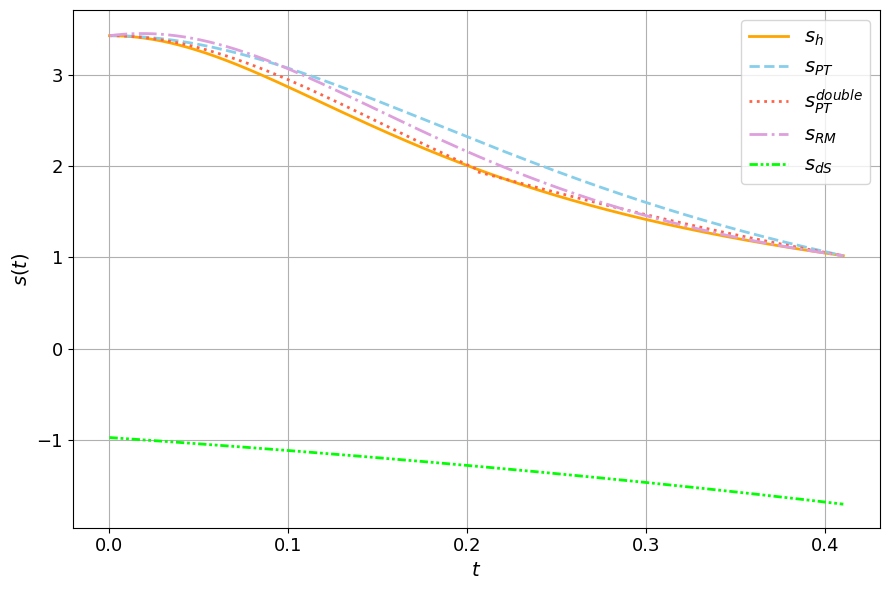} 
    \caption{Comparison of the exact effective mass in hybrid LQC, $s_h$ (solid orange), with the PT potential, $s_{PT}$ (dashed skyblue), and the three alternative mass terms: the double PT, $s_{PT}^{double}$ (dotted salmon), the RM potential, $s_{RM}$ (dot-dashed plum), and the de Sitter mass, $s_{dS}$ (dot-dot-dashed lime). The evolution of all masses is shown in the cosmological time $t$.}
    \label{Fig_masses}
\end{figure}

In this section we explore three alternative options to model the effective mass of the scalar perturbations near the quantum bounce in hybrid LQC: a double PT potential, a RM potential and a dS mass term. For this purpose, it is important to emphasize that we will keep in all cases the same value for the transition time $t_0$ as in the conventional PT case. For all these models, we can analytically obtain the mode solutions for the perturbations and the vacuum state selected by our criterion near the bounce. The three different masses are depicted in Fig. \ref{Fig_masses}, together with the PT potential and the exact hybrid LQC mass, $s_h$ (the latter computed by Eq. \eqref{LQC_mass}). \newline

\subsection{The double Pöschl-Teller potential}
\label{Subsec:double_PT_mass}

Instead of approximating the mass term $s_h$ by a PT potential, we want to derive a better approximation by composing two of these potentials, combined in the piecewise function

\begin{equation}\label{mass_double_PT}
s_h \approx s_{PT}^{double} =
\left\{
\begin{aligned}
s_{PT}^I &= \frac{U_1}{\cosh^2(\alpha_1 \eta)},
& 0 \le t \le t_0/2\\
s_{PT}^{II} &= \frac{U_1}{P^2 \cosh^2(Q\, \alpha_1 \eta)},
& t_0/2 < t \le t_0
\end{aligned}
\right. \, \, .
\end{equation}
In this way, we cover the initial period by two equally long intervals in cosmological time. In a sense, we are dividing the quantum era into two parts. Since the PT potential in each subinterval depends on two parameters (namely, $U_1$ and $\alpha_1$ in the first subinterval, and additionally $Q$ and $P$ in the second one), even if we require continuity of the effective mass, we have now an additional parameter at our disposal to improve the approximation obtained with a single PT potential. 

As in that case of a single PT potential, we demand that the approximated and the exact mass coincide at the bounce and at the final point of the quantum era, $t_0$. These conditions respectively determine the parameters $U_1$ and $Q$, the latter in terms of $\alpha_1$ and $P$. For $\alpha_1$, on the other hand, we simply demand that $s_{PT}^I$ matches $s_h$ at the transition time $t_0/2$. This condition, together with the two previous ones, the concavity property of the $1/\cosh^{2}$ function and the fact that $s_{PT}\geq s_h$ at $t_0/2$, can be shown to guarantee that our double PT potential improves the single PT approximation to the effective mass. In total, the three imposed conditions leave only one free parameter, $P$. In terms of it, we explicitly obtain 

\begin{equation}\label{parameters_double_PT}
  U_1 = \frac{8\pi \rho_c}{3}, \qquad Q = \frac{\operatorname{arccosh}(a_0^2/P)}{\alpha_1 \eta_0 }, \qquad \alpha_1= \frac{\operatorname{arccosh}(1 + 6\pi \rho_c t_0^2)^{1/3}}{\eta(t_0/2) } \, .
\end{equation}

Finally, for $P$ we choose a value that leads to a small relative error between $s_h$ and $s_{PT}^{double}$, and certainly smaller than the error for the single PT potential\footnote{We compute this error using Eq. \eqref{relative_error}.}. Concretely, we take $P=a^2(3t_0/10)$ (where the parentheses indicate the argument of the square scale factor), for which the error remains below $0.04$ throughout the entire quantum period for the quadratic inflaton potential. 

Although these elections provide the best approximation, there is a small discontinuity in the double PT mass at $t_0/2$. As commented above, it does not pose any obstacle when characterizing the evolution of the perturbations, since we impose continuity of the perturbative modes up to the first derivative at the transitions. 

We can use numerical methods to check and optimize the improvement of the double to single PT approximation to the effective mass. Actually, we have found a very close resemblance between the resulting numerical values and their analytic counterparts given above, which provides an additional consistency check for our construction.   

On the other hand, although the effective mass is described in both sections by a PT potential, these are two distinct functions, which implies different dynamics for the perturbations: same functional form, but different solutions. 

The mode solutions for the first subinterval of the quantum era ($0\leq t \leq t_0/2$) are given by

\begin{equation}\label{fisrt_PT_mode_solutions}
    \mu_k = \bar{M}_k \left [ z(1 - z) \right ]^{-ik/(2\alpha_1)} {}_{2}F_1 \left (c_1^k, c_2^k; c_3^k; z \right ) + \bar{N}_k  \left [\frac{z}{1 - z} \right]^{ik/(2\alpha_1)} {}_{2}F_1 \left (c_1^k - c_3^k + 1, c_2^k - c_3^k + 1; 2 - c_3^k; z \right ).
\end{equation}
This expression is analogous to Eq. \eqref{LQC_mass}, but here $ z = x(\alpha \rightarrow \alpha_1)$ and the parameters $c_i^k=b_i^k(\alpha \rightarrow \alpha_1, U_0 \rightarrow U_1)$ for $i=1,2,3$. In other words, for $z$ and $c_i^k$ we use the same functional expressions as previously for $x$ and $b_i^k$, but evaluated now at the new PT parameters, instead at those of the single potential approximation.

It is in this subinterval that we apply our criterion to choose the vacuum state, requiring that it provides an asymptotic Hamiltonian diagonalization in the ultraviolet. This selects the specific values $\bar{M}_k = 1 / \sqrt{2k}$ and $\bar{N}_k = 0$ of the integration constants in the mode solution, similarly to what happened in the single PT case. 

In the second subinterval, the mode solutions are determined as

\begin{equation}\label{second_PT_mode_solutions}
    \mu_k = \bar{F}_k \left [ \bar{x}(1 - \bar{x}) \right ]^{-ik/(2\alpha_2)} {}_{2}F_1 \left (d_1^k, d_2^k; d_3^k; \bar{x} \right ) + \bar{G}_k  \left [\frac{\bar{x}}{1 - \bar{x}} \right]^{ik/(2\alpha_2)} {}_{2}F_1 \left (d_1^k - d_3^k + 1, d_2^k - d_3^k + 1; 2 - d_3^k; \bar{x} \right ).
\end{equation}
In this case, $\bar{x}=x(\alpha \rightarrow \alpha_2)$ and $d_i^k=b_i^k(\alpha \rightarrow \alpha_2, U_0 \rightarrow U_2)$ for $i=1,2,3$. For convenience, we have introduced the notation $U_2 = U_1/P^2$ and $\alpha_2 = Q \, \alpha_1$ for the second PT potential. The integration constants $\bar{F}_k$ and $\bar{G}_k$ are selected by demanding continuity of the perturbative modes up to the first time derivative at $t_0/2$.

\subsection{The Rosen-Morse potential}
\label{Subsec:RM_mass}

The RM potential is given simply by the addition of a hyperbolic tangent term to a PT potential,

\begin{equation}\label{mass_RM}
    s_{RM} = \frac{V_0}{\cosh^2(\beta \eta  )} + V_1\tanh(\beta \eta ).
\end{equation}

The conditions to fix the parameters that best approximate the hybrid LQC mass $s_h$ for an RM potential are the same two imposed on the single PT potential (namely, the values at the beginning and at the end of the quantum era are equal to those of the exact mass), together with the demand that the second derivatives of both approximations to the effective mass coincide at the bounce. We then get

\begin{equation}\label{parameters_RM}
  V_0 = \frac{8\pi \rho_c}{3}, \qquad \beta = 4\sqrt{\pi \rho_c} \, , \qquad V_1 = \frac{V_0}{\tanh(\beta \eta_0)} \left (\frac{1}{a_0^{4}} - \frac{1}{\cosh^2(\beta \eta_0 )} \right ).
\end{equation}
This way, the relative error between $s_h$ and $s_{RM}$ grows at most up to $0.08$ for the whole quantum era for the quadratic inflaton potential. 

With this approximation, the perturbative modes evolve as 

\begin{equation}\label{RM_mode_solutions}
\begin{aligned}
      \mu_k = & \tilde{M}_k \, \tilde{x}^{-i\sqrt{k^2 - V_1} / (2\beta)} \left ( 1 - \tilde{x} \right)^{-i\sqrt{k^2 + V_1}/(2\beta)} {}_{2}F_1 \left (a_1^k, a_2^k; a_3^k; \tilde{x} \right ) \\
      & +  \, \tilde{N}_k \, \tilde{x}^{i\sqrt{k^2 - V_1}/ (2\beta)} \left (1 - \tilde{x} \right)^{-i\sqrt{k^2 + V_1} / (2\beta)} {}_{2}F_1 \left (a_1^k - a_3^k + 1, a_2^k - a_3^k + 1; 2 - a_3^k; \tilde{x} \right ),
\end{aligned}
\end{equation}
where $\tilde{x}=x(\alpha \rightarrow \beta)$ and the parameters of the hypergeometric function take the form

\begin{equation}\label{RM_hypergeometric_parameters}
    a_{1}^k = \frac{1}{2} \left (1 + \sqrt{1 + \frac{4V_0}{\beta^2}} \right ) - \frac{i}{2\beta} \left (\sqrt{k^2 + V_1} + \sqrt{k^2 - V_1} \right), \qquad a_{2}^k=a_{1}^k -  \sqrt{1 + \frac{4V_0}{\beta^2}},
\qquad a_3^k = 1 - \frac{i}{\beta}\sqrt{k^2 - V_1}.
\end{equation}

These expressions clearly evidence the asymmetric behaviour of the RM potential at the asymptotic limits of its argument. Although the mode solutions in the quantum era cover a considerably small interval in our analysis ($0\leq t \leq t_0$ in cosmological time), the differential equation \eqref{equation_MS} can be imposed at all times, beyond its domain of validity in the model considered in this work, with the commented asymmetry of the potential being then reflected in its general solution. 

In fact, it is also present in the complex frequency function selected by our criterion of vacuum choice, based on the asymptotic Hamiltonian diagonalization in the ultraviolet. We explicitly obtain 

\begin{equation}\label{RM_complex_freq_function}
   h_k(\tilde{x}) = -i\sqrt{k^2 - V_1} \, - i \left (\sqrt{k^2 + V_1} - \sqrt{k^2 - V_1} \right)\tilde{x} \, -2\beta \tilde{x}(1 - \tilde{x})\frac{\mathrm{d}}{\mathrm{d}\tilde{x}} \ln \left [ {}_{2}F_1 \left (a_1^k - a_3^k + 1, a_2^k - a_3^k + 1; 2 - a_3^k; \tilde{x} \right )\right ],
\end{equation}
which is well-defined for all $\tilde{x}$ and $k>0$. 

\begin{figure}[h]
    \centering
    \includegraphics[width=0.7\textwidth]{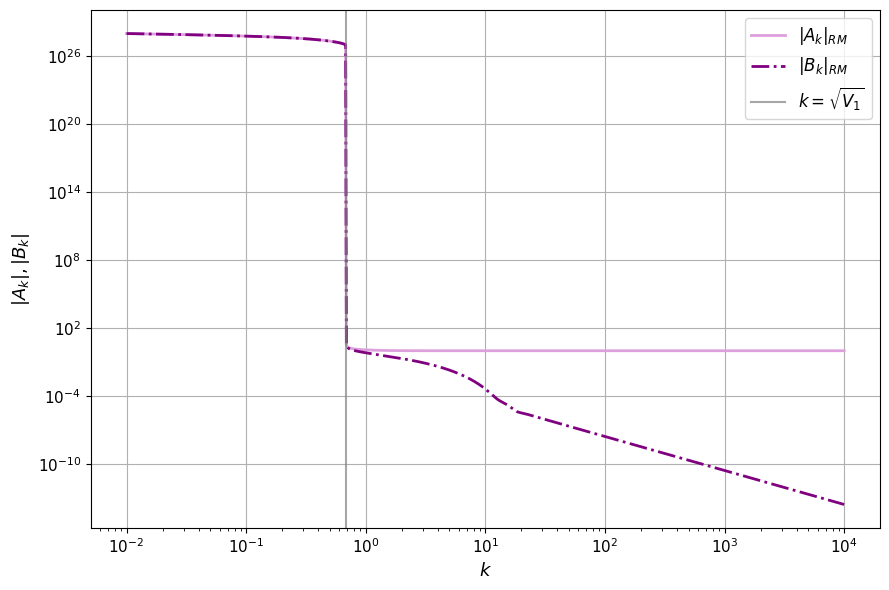}
        \caption{Variation with the wavenumber $k$ of the norms of the constants $A_k$ and $B_k$ for the RM potential, denoted by $|A_k|_{RM}$ and $|B_k|_{RM}$. These norms have been computed using Eqs. \eqref{RM_Ak} and \eqref{RM_Bk}, for $\gamma=0.2375$ and $t_0=0.41$. In solid grey, we display the critical scale $k=\sqrt{V_1}$, where $V_1$ is the parameter multiplying the $\tanh$ contribution in the RM potential.}
        \label{Fig_norms_RM}
\end{figure}

The vacuum solution is then determined by Eq. \eqref{NO-AHD_modes} for each mode provided that the imaginary part of the complex frequency function remains negative. However, this imaginary part vanishes when the wavenumber $k$ approaches $\sqrt{V_1}$ towards the ultraviolet sector. Therefore, in practice, our characterization of the vacuum state breaks down for $k\leq \sqrt{V_1}$. Moreover, as one could expect, the corresponding integration constants $A_k$ and $B_k$ of the mode solution in the slow-roll epoch display rapid changes around $k\approx \sqrt{V_1}$ that affect not only their phase, but also their norm. These changes cannot be absorbed by a genuine Bogoliubov transformation, in contrast to the situation found for spurious oscillations in the phases \cite{NM}. In Fig. \ref{Fig_norms_RM} we display the resulting plots. It should be noted here that the computation of the aforementioned norms for scales $k \leq \sqrt{V_1}$ is extremely sensitive to the precision of the numerical simulations, as we could expect for a numerically unstable problem. Nevertheless, we will later show (see Fig. \ref{Fig_PPS_RM}) that all this happens in the sector of wavenumbers where the PPS is already suppressed, a fact that trivializes the corresponding calculation of the spectrum. 

On the other hand, for $k>\sqrt{V_1}$ the results are well founded. The vacuum solution near the bounce is well defined and the NO-AHD solution at the end of inflation can be consistently determined. In particular, the norms of $A_k$ and $B_k$ do not oscillate too rapidly for these wavenumbers, nor display sudden changes in their variation (see Fig. \ref{Fig_norms_RM}). 
 
\subsection{The de Sitter case}
\label{Subsec:dS_mass}

In this scenario, we replace the LQC evolution of the background in the Planck regime by a phase of exact de Sitter dynamics. We follow the framework of delayed inflation developed in Ref. \cite{KM}, for the case of a kinetically dominated transition to the epoch of slow-roll. In our analysis, we do not intend to refine this evolution or track that of the corresponding cosmological constant that drives the expansion, rather we want to study whether the mass term for de Sitter expansion in the quantum era leads to a very different PPS adopting again the NO-AHD prescription. In fact, whenever needed, we will take the same initial values for the background variables at the beginning of the kinetically dominated epoch as in the conventional single PT model\footnote{With this choice, we get a value of $\rho_{dS} \approx 0.136\rho_c$ and a cosmological constant $\Lambda \approx 1.394$ in our Planck units.}. 

The scale factor during this dS epoch reads

\begin{equation}\label{scale_factor_dS}
    a(\eta) = \frac{a_0}{1 - a_0 H_0 (\eta - \eta_0)},
\end{equation}
where $a_0$ and $H_0$ are again the values of the scale factor and the Hubble parameter at the beginning of the kinetic dominance.

The time-dependent mass of the perturbations in dS is given by $-a''/a$, which yields

\begin{equation}\label{conformal_dS_mass}
    s_{dS} = -\frac{2(a_0H_0)^2}{\left [1 - a_0H_0(\eta - \eta_0) \right ]^2} \,.
\end{equation}

After computing the relation between proper and conformal times by $dt=a(\eta)d\eta $, we obtain 

\begin{equation}\label{proper_dS_mass}
   s_{dS} = - 2(a_0H_0)^2 e^{2H_0(t - t_0)}.
\end{equation}

In Fig. \ref{Fig_masses}, it can be seen that this mass term provides a really much worse approximation to the hybrid LQC mass $s_h$ than the other studied cases. This mass term leads to mode solutions of the form

\begin{equation}\label{dS_mode_solutions}
    \mu_k = \check{M}_k \frac{e^{ik\left (\eta - \eta_0 - 1/a_0H_0 \right)}}{\sqrt{2k}} \left [1 + \frac{i}{k\left (\eta - \eta_0 - 1/a_0H_0 \right)} \right ] + \check{N}_k \frac{e^{-ik\left (\eta - \eta_0 - 1/a_0H_0 \right)}}{\sqrt{2k}} \left [1 - \frac{i}{k\left (\eta - \eta_0 - 1/a_0H_0 \right)} \right ], 
\end{equation}
where $\check{M}_k$ and $\check{N}_k$ are the two corresponding constants. In this case, our criterion for the choice of a vacuum with asymptotic Hamiltonian diagonalization during the quantum epoch naturally selects the Bunch-Davies state for the perturbative modes, as one would expect for a dS regime, so that $\check{M}_k=0$ and $\check{N}_k=1$.

\section{Primordial power spectrum}
\label{Sec:PPS}

The PPS for scalar perturbations is defined as \cite{Baumann,Langlois}

\begin{equation}\label{PPS_formula}
    \mathcal{P}_{\mathcal{R}}= \frac{k^3}{2\pi^2} \left ( \frac{|\mu_k(\eta_f)|}{z(\eta_f)} \right ) ^2 ,
\end{equation}
where $z^2=a^2\epsilon_V/(4\pi)$ in the slow-roll regime. Approximately, the window of wavenumbers explored by observational missions is $2\times 10^{-4}\,\text{Mpc}^{-1}\leq k/a_{today} \leq 6 \times 10^{-1}\,\text{Mpc}^{-1}$ \cite{Planck_Constraints}. As mentioned in  Sec. \ref{Sec:Initial_framework}, given that we take an initial value of $a_B=1$ at the bounce, whereas it is the present value of the scale factor what is usually set to the unit for observational measurements, the corresponding window of observable modes has to be determined taking into account that $a_{today}=e^{n_T} a_B$, where $n_T$ is the total number of e-folds of expansion from the bounce to the present. For the case of phenomenological interest in which preinflationary effects have an impact on observable scales in a way that alleviates certain CMB tensions \cite{AG1,AM}, numerical studies in LQC indicate that the range $130\leq n_T\leq 143$ is compatible with these effects not being erased by an excessively large inflationary period \cite{AG1,AM}, and are also consistent with results found in recent analysis on the number of e-folds from the end of inflation until today \cite{ZSV}. With this in mind and converting inverse megaparsecs to inverse Planck units, for concreteness we will take the interval $k\in[10^{-2}, 10^4]$ in our numerical computations, which is potentially observable and captures the preinflationary effects we are interested in.

Thus, evaluating the PPS at a sufficiently large instant $\eta_f$ within or by the end of inflation for which these relevant modes have already frozen (which means that $k\ll a(\eta_f)H(\eta_f)$ for them), we can check from Eq. \eqref{conformal_time_sr} that it is possible to employ the asymptotic expansion of the Hankel functions for small arguments \cite{Abra} in solutions \eqref{sr_mode_solutions}. 

After some algebra, the PPS can then be computed to be \cite{GLMRV}

\begin{equation}\label{PPS_formula_sr}
    \mathcal{P}_{\mathcal{R}}= \left |A_k - B_k \right |^2 \times A_s \left (\frac{k}{k_*} \right ) ^{n_s-1}, \qquad A_s = \left (\frac{k_*}{k_f} \right)^{n_s - 1}\frac{H_f^2}{\pi \epsilon_V}, \qquad n_s = 4 - 2\nu,
\end{equation}
where $n_s$ is the spectral index, the scalar amplitude $A_s$ is determined as in Ref. \cite{GLMRV}, with the definition $k_f=(\eta_e - \eta_f)^{-1}$, and $k_*$ is the so-called pivot scale, an arbitrary reference scale at which the scalar amplitude is defined. We adopt the Planck convention for scalar perturbations and set $k_* = 0.05$, which is chosen to reduce parameter correlations in the analysis of observational data \cite{Planck_Constraints,Planck_parameters}. 

If the vacuum coincided with the Bunch-Davies state at the end of inflation, the above formula would yield the well-known quasi scale-invariant PPS. In our case, since the preinflationary epoch is relevant in the cosmological history and, in particular, in the determination of a natural vacuum, the Bunch-Davies state ceases to be a preferred choice. Any other choice of vacuum can be achieved as a Bogoliubov transformation of the Bunch-Davies state at the end of inflation, changing the corresponding (normalized) constants of the mode solution, $A_k$ and $B_k$. Thus, the PPS can be parametrized as the product of the standard quasi scale-invariant PPS and a preinflationary prefactor, given in the above formula by $\left |A_k - B_k \right |^2$, which accounts for the different choice of vacuum and, in turn, for the dynamics during the preinflationary era. 

\begin{figure}[h]
    \centering
    \includegraphics[width=0.7\textwidth]{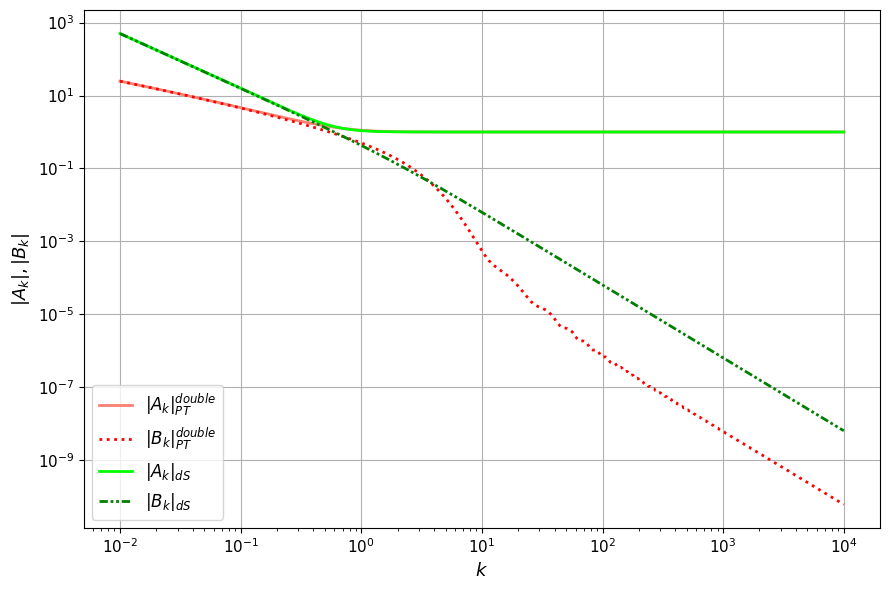}
        \caption{Variation with the wavenumber $k$ of the norms $|A_k|_{PT}^{double}$ (solid salmon), $|B_k|_{PT}^{double}$ (dotted red), $|A_k|_{dS}$ (solid lime), and $|B_k|_{dS}$ (dot-dot-dashed green), computed using Eqs. \eqref{double_PT_Ak}, \eqref{double_PT_Bk}, \eqref{dS_Ak} and \eqref{dS_Bk} respectively, for $\gamma=0.2375$ and $t_0=0.41$.}
        \label{Fig_norms_double_PT_and_dS}
\end{figure}

In the same direction, as briefly discussed in the Introduction, rather than considering the PPS of the state selected by our asymptotic Hamiltonian diagonalization in the quantum epoch, after evolving it through the kinetic domination and slow-roll regimes, we still have to apply a Bogoliubov transformation on it to reach a genuine NO-AHD vacuum at the end of inflation. This transformation removes spurious power oscillations caused by the lack of smoothness in our approximations. It consists of replacing $A_k \rightarrow |A_k|$, $B_k \rightarrow |B_k|$, provided that the norms of $A_k$ and $B_k$ do not oscillate rapidly. This can be seen to be satisfied for the RM potential in Fig. \ref{Fig_norms_RM}, for the sector of wavenumbers in which the NO-AHD state is well-defined, and for the double PT and the dS masses in Fig. \ref{Fig_norms_double_PT_and_dS}. This transformation is equivalent to selecting the envelope of the minima of the fast oscillations in the PPS \cite{NM}. This way, the PPS reads \cite{GLMRV}

\begin{equation}\label{PPS_transformed}
    \mathcal{P}_{\mathcal{R}}= \Big (|A_k| - |B_k| \Big )^2 \times  \mathcal{P}_{\mathcal{R}}^{\Lambda CDM}, \qquad  \mathcal{P}_{\mathcal{R}}^{\Lambda CDM}= A_s \left (\frac{k}{k_*} \right ) ^{n_s-1}.
\end{equation}

Therefore, once the slow-roll constants $A_k$ and $B_k$ for our choice of vacuum are determined via the continuity requirements on the mode solutions at the matching points, we obtain the PPS in the NO-AHD prescription. To simplify the expression of the Bogoliubov prefactor, we will adopt the same simplifications explained in Ref. \cite{GLMRV}, which are based on the comparison of the different scales involved in the cosmological evolution. Essentially, all wavenumber scales related to the onset and duration of slow-roll inflation are much smaller than those corresponding to the quantum epoch, allowing for suitable asymptotic expansions in our mode solutions.

To be more specific, before the aforementioned simplifications, the expression for the PPS of the NO-AHD vacuum depends, apart from the standard parameters of the $\Lambda$CDM model, on four additional parameters, which are the suppression scale $k_0=a_0H_0$ and the duration of the three stages of the evolution, each computed in the form of different parameters. In addition, we recall that, when one adopts the convention that the present scale factor is one, all lengths must be rescaled by the total number of e-folds of expansion since the bounce. For the window of wavenumbers of phenomenological interest \cite{AM,AG1,NM,GLMRV}, the commented simplifications remove from the expression of the PPS the two parameters related to the duration of the kinetic domination and the slow-roll phase by applying the asymptotic expansion of the Hankel functions for large arguments \cite{Abra}. Thus, the Bogoliubov factor only depends on two extra parameters: the scale factor at the end of the quantum era $a_0$ (or, equivalently, the number of e-folds during this quantum epoch) and the aforementioned suppression scale $k_0$, which contains the total number of e-folds as we have explained. This way, the parameters related to the physics during the Planck regime have the dominant impact on the shape of the PPS. Another important remark is that $a_0$ and $k_0$ determine the critical density of LQC (with all of our conventions) as $\rho_c=3k_0^2 a_0^4/(8\pi)$.

Following this line of reasoning, our goal is to parametrize the PPS for each of the three cases considered in the previous section in terms of $a_0$ and $k_0$, and then compare the results to the spectrum obtained with the single PT potential. In the rest of this section, we present the expressions of the slow-roll coefficients $A_k$ and $B_k$ of our vacuum state for each of the three studied mass terms. The plots of the resulting PPS are displayed in the following section. 

\subsection{Primordial power spectrum for the double Pöschl-Teller potential}
\label{Subsec:PPS_double_PT}

In this case, the simplified norms of $A_k$ and $B_k$ are found to be

\begin{equation}\label{double_PT_Ak}
\begin{aligned}
    \left|A_k\right|^{double}_{PT}  =   \frac{1}{4}\sqrt{\frac{\pi k}{k_0}} \, \Biggl |  & \, \frac{{}_{2}F_1 \left (c_1^k, c_2^k; c_3^k; z_m \right )}{[h(a_0)]^{ik/(3\alpha_1)}}  \left (\frac{\mathrm{d}}{\mathrm{d}\bar{x}}\ln \frac{{}_{2}F_1 \left (d_1^k - d_3^k + 1 , d_2^k - d_3^k + 1; 2 -  d_3^k; \bar{x} \right )\bar{x}^{ik/(Q\alpha_1)}}{{}_{2}F_1 \left (d_1^k, d_2^k; d_3^k; \bar{x} \right )} \right)^{-1}_{\bar{x}_m} \, H_0^{(1)}\left (\frac{k}{2k_0} \right) \Biggr | \\
      & \times  \Biggl |   \frac{{}_{2}F_1 \left (d_1^k, d_2^k; d_3^k; \bar{x}_0 \right )}{{}_{2}F_1 \left (d_1^k, d_2^k; d_3^k; \bar{x}_m \right )}\, \left ( \frac{2a_0^2\sqrt{f(a_0)}}{P}\right )^{ik/(Q\alpha_1)}  \, \text{B}_{1} \, \text{B}_{2} \\
      & -    \frac{{}_{2}F_1 \left (d_1^k-d_3^k +1, d_2^k- d_3^k +1; 2-d_3^k; \bar{x}_0 \right )}{{}_{2}F_1 \left (d_1^k-d_3^k +1, d_2^k- d_3^k +1; 2-d_3^k; \bar{x}_m \right )}\,\big [g(a_0 )\big ]^{ik/(2Q\alpha_1)} \, \text{B}_{3} \, \text{B}_{4}    \Biggr |,
\end{aligned}
\end{equation}

\begin{equation}\label{double_PT_Bk}
\begin{aligned}
    \left|B_k\right|^{double}_{PT}  = \left|A_k\right|^{double}_{PT} \Big ((1) \leftrightarrow (2) \Big )_{0,1} \, . 
\end{aligned}
\end{equation}
The notation $\big  ((1) \leftrightarrow (2) \big )_{0,1}$ means that the norm of $B_k$ is the same as $A_k$ but swapping the kind of the Hankel functions for both orders $0$ and $1$. In addition, we have used the same notation as in Sec. \ref{Subsec:double_PT_mass}, the subindex $\bar{x}_m$ in the big parentheses stands for evaluation at this value of $\bar{x}$, and we have defined $\bar{x}_0=\bar{x}(\eta_0)$, $\bar{x}_m=\bar{x}(\eta_m)$, and $z_m = z(\eta_m)$, where $\eta_m = \eta(t_0/2)$. Note that it is possible to express all these last quantities as functions of $a_0$. Besides, we have introduced the $a_0$-dependent functions

\begin{equation}\label{functions_a0}
\begin{aligned}
    &  f(a_0)= \left [2\cosh\left (Q\operatorname{arccosh} \left ( \frac{3+a_0^6}{4}\right)^{1/3} \right ) \right ]^{-2} ,\\
    & g(a_0) =  \left [\frac{a_0^2}{P}+ \sqrt{\left (\frac{a_0^2}{P}\right)^2 - 1} \, \right ]^{2-C(a_0)}, \qquad C(a_0) = \frac{{}_{2}F_1  \left (\frac{1}{6}, \frac{1}{2}; \frac{3}{2}; \frac{1-a_0^6}{4} \right )}{{}_{2}F_1  \left (\frac{1}{6}, \frac{1}{2}; \frac{3}{2}; 1-a_0^6 \right )}.
\end{aligned}
\end{equation}
Lastly, the factors $B_i$ ($i=1,2,3,4$) are given by

\begin{equation}\label{B_1}
\begin{aligned}
    \text{B}_1 =   \left (\frac{\mathrm{d}}{\mathrm{d}z}\ln \frac{{}_{2}F_1 \left (c_1^k, c_2^k; c_3^k; z \right )}{[z(1-z)]^{ik/(2\alpha_1)}} \right )_{z_m} \, \frac{\big [h(a_0) \big ]^{2/3}}{f(a_0)Q} - \left (\frac{\mathrm{d}}{\mathrm{d}\bar{x}}\ln \frac{{}_{2}F_1 \left (d_1^k- d_3^k + 1, d_2^k - d_3^k + 1; 2- d_3^k; \bar{x} \right )\bar{x}^{ik/(2Q\alpha_1)}}{(1-\bar{x})^{ik/(2Q\alpha_1)}} \right )_{\bar{x}_m} \, ,  
\end{aligned}
\end{equation}

\begin{equation}\label{B_2}
\begin{aligned}
   \text{B}_2 =  \left (\frac{\mathrm{d}}{\mathrm{d}\bar{x}}\ln \frac{{}_{2}F_1 \left (d_1^k, d_2^k; d_3^k; \bar{x} \right )}{[\bar{x}(1-\bar{x})]^{ik/(2Q\alpha_1)}} \right )_{\bar{x}_0} \, \frac{Q\alpha_1}{2k}\left (\frac{P}{a_0^2} \right)^2   - \frac{k_0}{k} + \frac{H_1^{(1)}\left (\frac{k}{2k_0} \right)}{H_0^{(1)}\left (\frac{k}{2k_0} \right)} \, ,
\end{aligned}
\end{equation}

\begin{equation}\label{B_3}
\begin{aligned}
    \text{B}_3 = \text{B}_1 \left ( \frac{{}_{2}F_1  \left (d_1^k- d_3^k + 1, d_2^k - d_3^k + 1; 2- d_3^k; \bar{x} \right )\bar{x}^{ik/(2Q\alpha_1)}}{(1-\bar{x})^{ik/(2Q\alpha_1)}}  \rightarrow \frac{{}_{2}F_1 \left (d_1^k, d_2^k; d_3^k; \bar{x} \right )}{ [\bar{x}(1-\bar{x})]^{ik/(2Q\alpha_1)}}  \right )  ,  
\end{aligned}
\end{equation}

\begin{equation}\label{B_4}
\begin{aligned}
    \text{B}_4 = \text{B}_2 \left ( \frac{{}_{2}F_1 \left (d_1^k, d_2^k; d_3^k; \bar{x} \right )}{ [\bar{x}(1-\bar{x})]^{ik/(2Q\alpha_1)}} \rightarrow  \frac{{}_{2}F_1  \left (d_1^k- d_3^k + 1, d_2^k - d_3^k + 1; 2- d_3^k; \bar{x} \right )\bar{x}^{ik/(2Q\alpha_1)}}{(1-\bar{x})^{ik/(2Q\alpha_1)}}  \right )  ,
\end{aligned}
\end{equation}
where the arrows in the two last equations indicate the corresponding replacement in the expressions of $B_1$ and $B_2$, and $h(a_0) =1/[2(3 + a_0^6)]$. 

\subsection{Primordial power spectrum for the Rosen-Morse potential}
\label{Subsec:PPS_RM}

Taking into account the commented subtleties in determining the NO-AHD vacuum state for an RM mass term, we directly compute the simplified norms of $A_k$ and $B_k$ employing the modes \eqref{NO-AHD_modes} and the complex frequency function given in Eq. \eqref{RM_complex_freq_function}. This yields

\begin{equation}\label{RM_Ak}
\begin{aligned}
    |A_k|_{RM}  = \frac{k}{2}\sqrt{\frac{\pi}{2k_0}} \left |H_0^{(1)}\left (\frac{k}{2k_0} \right) \left [\frac{\mu_k'(\eta_0)}{k} - \left (\frac{k_0}{k} - \frac{H_1^{(1)}\left(\frac{k}{2k_0}\right)}{H_0^{(1)}\left(\frac{k}{2k_0}\right)} \right)\mu_k(\eta_0) \right ] \right |,
\end{aligned}
\end{equation}

\begin{equation}\label{RM_Bk}
\begin{aligned}
    |B_k|_{RM}  = |A_k|_{RM} \Big ((1) \leftrightarrow (2) \Big )_{0,1} \, ,
\end{aligned}
\end{equation}
where we have used a notation similar to that of the previous subsection.

\subsection{Primordial power spectrum for the de Sitter case}
\label{Subsec:PPS_dS}

In this model, the resulting simplified norms of $A_k$ and $B_k$ can be shown to take the form

\begin{eqnarray}\label{dS_Ak}
    |A_k|_{dS}  &=& \frac{1}{4}\sqrt{\frac{\pi k}{k_0}} \, \left |H_0^{(1)}\left (\frac{k}{2k_0} \right) + \left (i-\frac{k_0}{k} \right)H_1^{(1)}\left(\frac{k}{2k_0}\right) \right |,\\
\label{dS_Bk}
    |B_k|_{dS}  &=& |A_k|_{dS} \Big ((1) \leftrightarrow (2) \Big )_{0,1} \, ., 
\end{eqnarray}

using the same kind of notation as above.

\section{Discussion}
\label{Sec:Discussion}

In our discussion of the changes in the PPS compared to the standard result of $\Lambda$CDM cosmology, it is very convenient to employ the parametrization given in Eq. \eqref{PPS_transformed}. It accurately reflects the dependence of the PPS on the vacuum state of the perturbations, which, in the case of the NO-AHD prescription, is optimally adapted to the background evolution that dictates the dynamical mass entering the propagation equations of the scalar perturbations. In the literature, this mass has often been approximated by a PT potential. In order to investigate the influence of this approximation in the form deduced for the PPS and check the validity of the results obtained with it, we have considered other different approximations and preinflationary scenarios and studied whether they lead or not to a significantly different PPS. This way, we have been able to test the robustness of the predictions of the hybrid LQC model (with the choice of the NO-AHD prescription) for the PPS. 

As illustrated in Ref. \cite{GLMRV}, in this framework the preinflationary factor only depends in a relevant manner on two additional parameters compared to the $\Lambda$CDM model. These are the scale factor at the end of the quantum era, $a_0$, and the suppression scale, $k_0$. The former is related to the number of e-folds during the quantum epoch, whereas the latter is also connected to the total number of e-folds of expansion, relating the characteristic scale at the end of the quantum epoch to the length scale of the present Universe. For the three alternative scenarios that we have considered, we adopt the same values for these two parameters, equal to those given in Ref. \cite{GLMRV} for the single PT potential. In particular, this implies that all these cases have the same scale of suppression (except for minor subdominant details). Therefore, the only visible significant change can be the slope of the PPS.

\begin{figure}[htbp]
    \centering

    \begin{minipage}[t]{0.49\textwidth}
        \centering
        \includegraphics[width=\linewidth]{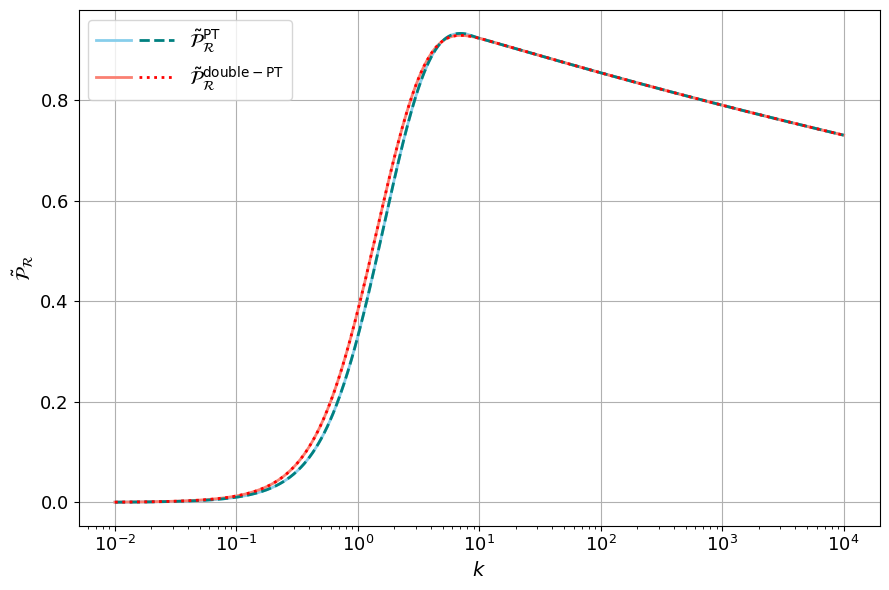}
        \caption{Comparison of the PPS for the PT and the double PT potentials. For the PT potential, the solid skyblue line is the result using the exact formula of the constants $A_k$ and $B_k$, whereas the dashed teal line is the result employing the simplified version of this formula, as explained in the text. For the double PT potential, the solid salmon line is the result for the exact formula and the dotted red line is the result employing its simplified version. We have taken $\gamma=0.2375$ and $t_0=0.41$.}
        \label{Fig_PPS_double_PT}
    \end{minipage}
   \hfill
	\begin{minipage}[t]{0.49\textwidth}
        \centering
        \includegraphics[width=\linewidth]{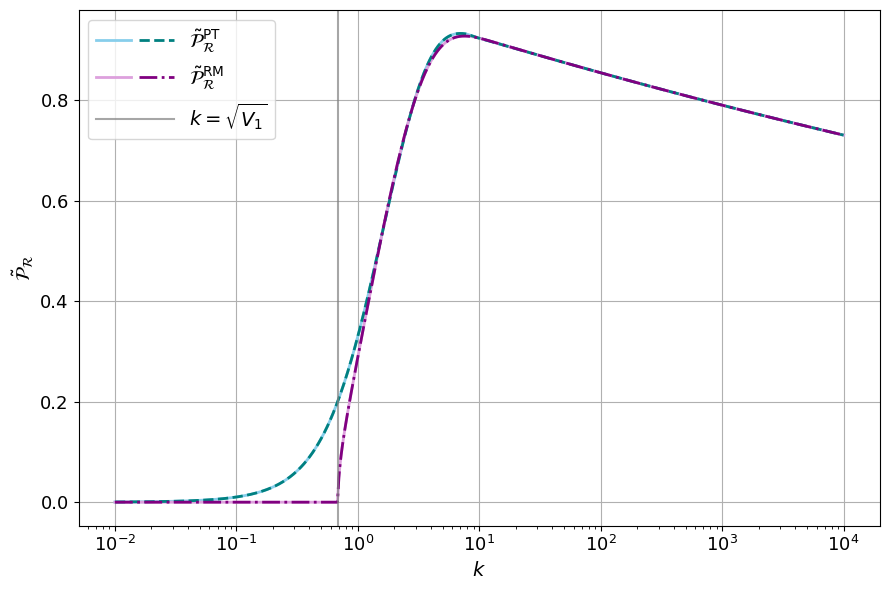}
        \caption{Comparison of the PPS for the PT and the RM potentials. For the PT potential, the solid skyblue line is the result using the exact formula of the constants $A_k$ and $B_k$, whereas the dashed teal line is the result employing the simplified version of this formula, as explained in the text. For the RM potential, the solid plum line is the result for the exact formula and the dot-dashed purple line is the result employing its simplified version. We have taken $\gamma=0.2375$ and $t_0=0.41$. In solid grey, we mark the critical scale $k=\sqrt{V_1}$.}
        \label{Fig_PPS_RM}
    \end{minipage}
\end{figure}

\begin{figure}[h]
    \centering
    \includegraphics[width=0.7\textwidth]{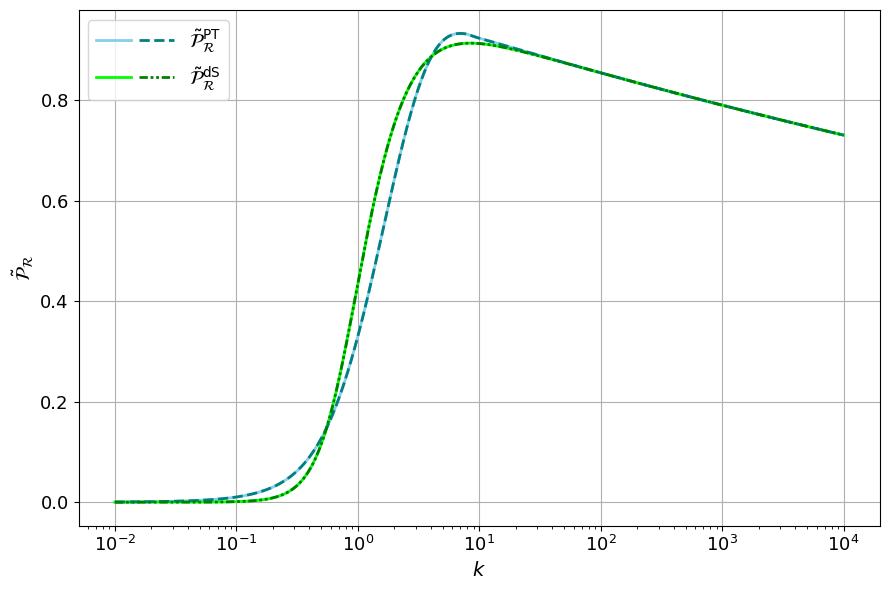}
        \caption{Comparison of the PPS for the PT potential and the dS case. For the PT potential, the solid skyblue line is the result using the exact formula of the constants $A_k$ and $B_k$, whereas the dashed teal line is the result employing the simplified version of this formula, as explained in the text. For the dS case, the solid lime line is the result for the exact formula and the dot-dot-dashed green line is the result employing its simplified version. We have taken the same values of $a_0$ and $k_0$ as in the considered models in hybrid LQC, obtained for $\gamma=0.2375$ and $t_0=0.41$.}
        \label{Fig_PPS_dS}
\end{figure}

This is confirmed by our plots of the PPS in Figs. \ref{Fig_PPS_double_PT}, \ref{Fig_PPS_RM}, and \ref{Fig_PPS_dS} for the double PT potential, the RM potential, and the dS case, respectively. In these plots, we have displayed the resulting PPS after the aforementioned simplification process, as well as its non-simplified counterpart, which can be directly computed numerically. We have also plotted the PPS for the single PT potential for comparison. For this last model, expressions can be found in Ref. \cite{GLMRV}.

The main feature that we observe in the spectrum of the NO-AHD vacuum of all the considered cases is the power suppression in the infrared region \cite{Linde,ASr,Enough_inf1}, at wavenumber scales that are indeed determined by the characteristic scale of the quantum preinflationary epoch (up to rescaling by the number of accumulated e-folds). Details on the specific scale at which this suppression occurs slightly depend on the particular model, but the sensitivity that we find is certainly small. In contrast, a feature that is more clearly model dependent is the steepness of the suppression, which differs in the considered scenarios.  

In total, given the choice of vacuum state for the perturbations based on the NO-AHD proposal \cite{NMT}, the modifications with respect to the PPS of a Bunch-Davies state are mainly encoded in the suppression scale and the slope of the PPS at it. The power suppression appears at a scale that can be estimated from the characteristic scale of the background at the end of the quantum epoch and the total number of e-folds accumulated in the cosmological history. The slope of the suppression, however, has a more intrincate dependency that accounts for other details of the physics felt by the perturbations near the bounce (as evidenced by its changes under variations of the mass term).

The PPS that most closely resembles that of the PT potential corresponds to the double PT approximation, with a tilt slightly less pronounced than in the PT case. On the contrary, the spectrum that most differs from its PT counterpart is provided by the dS case, with a considerably more inclined curve for $0.5 \lesssim k \lesssim 2$. This was expected given that this is the model in which the mass term differs the most from the hybrid LQC value. On the other hand, the RM curve is practically indistinguishable from the PT one until it reaches the surroundings of the critical scale ($k=\sqrt{V_1}$), where the power is suppressed more sharply. For smaller scales, the validity of the analysis breaks down in this case and the power is totally suppressed.  However, this problem does not affect much the physical results, because in that region of scales the PPS is severely suppressed also in the single PT model.

Our calculation of the PPS has been directly attained from Eq. \eqref{PPS_transformed}, with the norms of the slow-roll constants given in the previous section. As mentioned in Sec. \ref{Sec:Initial_framework}, the window of wavenumbers has been chosen by taking into account the shift with respect to the scales explored by observational missions. With this in mind and focusing on the scales that better capture the preinflationary effects we are interested in, we have taken the interval $k\in[10^{-2}, 10^4]$ for our numerical computations. We can check that this interval indeed satisfies the condition to implement the simplifications on the expressions of the constants $A_k$ and $B_k$ explained above. 

As an overall result, the three PPS that we have obtained are very similar to the one computed with the PT potential. Given the current resolution in observational missions, it is highly unlikely that it will be possible to tell them apart. Therefore, we can conclude that the predictions for the PPS of the NO-AHD vacuum in hybrid LQC are not significantly sensitive to the approximation employed for the effective mass of the scalar perturbations near the bounce. 

Furthermore, we expect this conclusion to be valid beyond the restrictions we have adopted in this work for concreteness. For example, we expect that for other inflaton potentials, tensor perturbations, or the dressed metric approach to LQC, this result will essentially be the same. In the scenarios studied, the potential contribution to the inflaton energy density in the vicinity of the bounce is irrelevant, so the suppression and slope of the PPS, which are primarily determined by the preinflationary dynamics of the quantum period in our NO-AHD prescription, will remain invariant under changes in the inflaton potential. Secondly, the difference in the effective mass of scalar and tensor perturbations is practically negligible in regions in which the potential energy is; therefore, the conclusion we have obtained holds for tensor perturbations too. Lastly, in the dressed-metric strategy, the effective mass for the perturbations can also be described by a PT potential in the quantum regime after the bounce. The resulting PPS for the NO-AHD vacuum shares similar aspects to its hybrid counterpart (with differences in the steepness of the spectrum and the value of the suppression scale \cite{GLMRV}), which makes it more than plausible that our results about the robustness of the adopted approximation apply to the dressed-metric approach too.

Since our results indicate that, for our choice of vacuum, the predicted PPS is not significantly dependent on the particular approximation adopted for the effective mass near the bounce, provided that the characteristic scale associated with the quantum era is correctly reproduced, an interesting question for further research is to quantify how much this scale is affected by details of the background quantum physics around this bounce and by different kinds of quantization ambiguities. In addition, it might be interesting to discuss other observables in cosmology with a different functional dependence on the total number of e-folds from the bounce until today and on the scale at the end of the bounce period, to break the degeneracy that we have found in the dependence of the power suppression scale in their product.

\acknowledgments

The authors are thankful to O. Friedrich and K. Langer for helpful conversations. They are also very thankful to B. Elizaga Navascu\'es for discussing fundamental ideas for this work, and to A. Vicente-Becerril and J. Y\'ebana-Carrilero for conversations and help with numerical calculations. A.G. would also like to thank Ryo Namba for interesting discussions. This work was partially supported by MCIU/AEI/10.13019/501100011033 and FSE+ under the Grant No. PID2023-149018NB-C41. The Spanish MCIU also supports A.G. under the FPU Grant No. FPU24/01521.




\end{document}